\documentclass[a4paper,preprint]{aastex}
\usepackage{graphicx}

\newcommand{\ri}{\ensuremath{\rho_{\rm I}}}
\newcommand{\unts}[1]{\ensuremath{\,\mathrm{#1}}}
\newcommand{\HI}{H\,\textsc{i}}

\newcommand{\dem}{DEM L\,316}

\begin{document}

\title{Remnants from Gamma-Ray Bursts}

\author{Shai Ayal and Tsvi Piran}

\affil{Racah Institute of Physics, Hebrew University,
  Jerusalem 91904, Israel} 

\begin{abstract}
  We model the intermediate time evolution of a ``jetted'' gamma-ray
  burst by two blobs of matter colliding with the interstellar medium.
  We follow the hydrodynamical evolution of this system numerically
  and calculate the bremsstrahlung and synchrotron images of the
  remnant. We find that for a burst energy of $10^{51}\unts{erg}$ the
  remnant becomes spherical after $\sim 5000$ years when it collects
  $\sim 50M_\odot$ of interstellar mass. This result is independent
  of the exact details of the GRB, such as the opening angle. After
  this time a gamma-ray burst remnant has an expanding sphere
  morphology. The similarity to a supernova remnant makes it difficult
  distinguish between the two at this stage.  The expected number of
  non-spherical gamma-ray burst remnants is $\sim0.05$ per galaxy for a
  beaming factor of $0.01$ and a burst energy of $10^{51}\unts{erg}$.
  Our results suggest that that the double-shell object \dem~is not a
  GRB remnant.
\end{abstract}

\section{Introduction}
\label{sec:Intro}

If a $\gamma$-ray burst (GRB) originate within a galactic disk then
the large deposition of energy will result in a blast wave whose
initial phase produces the afterglow. The late phase of the blast wave
evolution would result, as noted by \cite{chevalier74} in the context
of supernova remnants (SNRs), in a cool expanding \HI~shell. The shell
will remain distinct from its surrounding until it has slowed down to
a velocity of $\approx 10\unts{km\,s^{-1}}$. This would happen within
$2.3E_{51}^{0.32}10^6\unts{yr}$ where $E_{51}$ is the initial energy
in units of $10^{51}\unts{erg}$ \citep{loeb98}.  The current rate of
GRBs is one per $\sim10^7\unts{yr}$ per galaxy \citep{schmidt99}.
This leads us to expect a few remnants per galaxy at any given time.
However SN are $10^5$ times more frequent and it would be difficult to
distinguish between a SNR and a GRB remnant.  Today there is mounting
evidence that some GRBs are beamed
\citep{halpern99,harrison99,kulkarni99,sari99a,kuulkers00,salmonson00}.
Beamed GRBs will illuminate only a fraction $f_b$ of the sky, and
their rate should be higher by a factor of $f_b^{-1}$. With $f_b\sim
0.01$ we would expect a hundred GRB remnants per galaxy. At an early
stage the morphology of a ``jetted'' GRB remnant would be very
different from a spherical explosion. This can be used to identify
those remnants. We study this phase here.

Both beamed GRBs and SN deposit a comparable ($\sim
10^{51}\unts{erg}$) energy into the ISM. In both cases the evolution
is expected to be similar since both are in the Sedov \citep{Sedov59}
regime where all the energy is the initial explosion energy and all
the mass is in the surrounding ISM. A key distinguishing feature
unique to GRB remnants could be their beamed nature which we expect
would lead to a distinct double shell morphology at intermediate times
(the late time behavior of the GRB remnant is expected to be spherical
in any case). In order to establish how many \HI~shells are GRB
remnants we must find out the expected morphology of GRB remnants and
how long they remain non-spherical, and distinguishable from SNRs.

Expanding \HI~shells have been found in many spiral galaxies
\citep{tenorio-tagle88}. The interior of these shells is relatively
empty and their current expansion velocity is in the order of tens of
Km s$^{-1}$. Models for the origin of these \HI~supershels involve a
large number of spatially correlated supernova events
\citep{heiles79,heiles84}, infall of massive gas clouds on to the
galactic plane \citep{tenorio-tagle88} and flaring of radio lobes
formed by jets ejected from the galactic nucleus during an active
phase \citep{gopal-krishna00}. It has been previously noted
\citep{efremov98,loeb98} that a subset of these \HI~supershells may be
the late signature left by GRBs on the interstellar medium (ISM).
Establishing how many of the \HI~shells are GRB remnants would make it
possible to directly estimate the local rate of GRBs, determine
$\epsilon$, the efficiency of converting the explosion energy into
$\gamma$-rays, and the beaming factor $f_b$ \citep{loeb98}.

We model the intermediate evolution of a beamed GRB by two blobs of
dense material moving into the ISM in opposite directions and we
follow numerically their hydrodynamical evolution. Our results can be
rescaled to fit a variety of initial energies. We find that at a time
of $\sim 5000\unts{yr}$, when the ratio $Mc^2/E_0$ between the
accumulated mass, $M$, and the initial GRB energy,$E_0$, reaches $\sim
9.6\times10^4$ the remnant becomes spherical, similar in shape to a
SNR. This value is independent of the exact initial details of our
model such as the opening angle, the velocity and the morphology. We
compare our results to the \HI~shells \dem~\citep{Williamsetal97}
previously classified as two colliding SNRs. As the accumulated mass
there is much larger it is most likely not a GRB remnant.

We describe our model and the numerical methods in section 2 . In
section 3 we describe our results. Discussion and summary including a
comparison with \dem~are given in section 4.

\section{The Simulation}
\label{sec:physit}
\subsection{Model}

We study the intermediate evolution of the relativistic ejecta that
caused a GRB. We assume that this matter was ejected in the form of
two ultra relativistic blobs moving in opposite directions with a bulk
Lorenz factor $\Gamma\gtrsim 100$. The emission from internal
collisions in the blobs during an early stage comprises the GRB. Late
external shocks caused by collisions with circumstellar matter produce
the afterglow. The matter slows down during this interaction and its
bulk Lorenz factor $\Gamma$, decreases. The ejecta stays collimated
only until $\Gamma$ drops below $\sim 1/\theta_0$, at approximately
$3.4(E_{51,\rm{iso}}/\ri)^{1/3}(\theta_0/0.1)^{8/3}\unts{hr}$ after
the GRB, when it has accumulated
$10^{-5}E_{51,\rm{iso}}(\theta_0/0.1)^2\unts{M_\odot}$ of ISM mass
\citep{sari99} where $\theta_0$ is the initial angular width,
$E_{51,{\rm iso}}$ the isotropic energy in units of
$10^{51}\unts{erg}$ and $\ri$ is the surrounding ISM density in units
of $10^{-24}\unts{gm\,cm^{-3}}$. Note that in the rest of the paper
$E_{51}$ denotes the {\em actual} energy in units of
$10^{51}\unts{erg}$. At this time the matter will start to expand
sideways causing, for an adiabatic evolution, an exponential slowing
down \citep{rhoads97}. The ejecta continues to expand sideways at an
almost constant radial distance from the source $R_0\sim
0.3E_{51}^{1/3}\ri^{-1/3}\unts{pc}$ until it becomes non-relativistic.
At this stage, we begin our simulation.

Without a detailed numerical modeling of the relativistic phase of the
ejecta we have only an approximate description of how to construct the
initial conditions. We expect the angular width of the ejecta to be
$\sim 1\unts{rad}$. Additionally we are constrained by Newtonian
energy conservation which yields a relation between $R_0$, $\Gamma$
and $E_0$:
\begin{equation}
R_0 \sim 0.3E_{51}^{1/3}\ri^{-1/3}(v_0/c)^{-2/3}\unts{pc}\,.\label{eq:const}
\end{equation}
The exact shape and energy distribution of the ejecta are uncertain.
This may influence the numerical coefficient in Eq. (\ref{eq:const}).
However we show that the intermediate and late evolution of the ejecta
are insensitive to these uncertainties in the initial conditions. This
is due to the fact that since we are in the Sedov regime, the mass is
dominated by the ``external'' ISM gas which washes out any variations
in the initial conditions of the ejecta. This same mechanism will
smear out any non-spherical features in a SN explosion, producing
spherical SNRs.

We realize these initial conditions by two identical blobs moving in
opposing directions into the ISM. The computational space is a
cylinder in which the blobs move along the symmetry axis. Both the
blobs and the ISM are modeled by a cold $\gamma=5/3$ ideal gas. The
blobs are denser than the ISM. We consider various initial densities,
angular widths and shapes of the blobs in order to make sure that
these are indeed unimportant in determining the final morphology. In
all runs $v_0\sim c/3$. To evolve these initial conditions we use a
hydrodynamic code, neglecting radiative effects.

We note that we have also used the post Newtonian version of our code
with the same initial blob velocity of $c/3$. We found that the blobs
slow down rapidly and the results were similar to the Newtonian
results.  Within a few timesteps we have a Newtonian system. In
essence this is equivalent to slightly changing the initial shape of
the blobs which, as noted before, is insignificant due to the mass
dominance of the ISM.

\subsection{The Numerical Method}
\label{sec:NumMeth}
The code we use is based on the Newtonian version of the smooth
particle hydrodynamics (SPH) code introduced in \cite{sphpn}. The code
was adapted for the specific problem at hand. The three
computationally important features of the problem are the negligible
role of gravity, the large amount of stationary gas representing the
ISM, and the symmetry.  The fact that gravity is unimportant has made
the equations very simple and greatly increased the speed of the code.
Implementing single particle time steps allowed us to put in a large
volume of stationary gas without making a big impact on the
computational time.  This way most of the computational effort is
invested in the ``important'' interacting particles. Typically only
0.1\% of the particles actually move during each timestep.

The initial conditions are set up so that the blobs are on the $z$
axis with a velocity along the $z$ axis. With these initial conditions
there is a rotational symmetry about the $z$ axis and a reflection
symmetry about the $x-y$ plane. We implement the reflection symmetry
exactly and the rotation symmetry only approximately. The
implementation, similar to \cite{larry93}, consists of considering
only $1/8$ (one quadrant) of the computational volume and adding
reflecting boundary conditions on the three inner boundaries defined
by the $x-y$, $x-z$, and $y-z$ planes. We implement no outer boundary
condition since the surrounding gas is very cold and there is almost
no pressure so the outer particles gain only a negligible velocity
throughout the course of the simulation. The quadrant we evolve
consists of 62,800 SPH particles of which only $\sim 10$ represent the
initial blob, again stressing the total mass dominance of the ISM.

The reflecting boundary conditions are implemented using
pseudo-particles. At the beginning of each timesteps, all particles
which intersect one of the boundaries (In SPH each particle has a
finite size called the smoothing length) are reflected about this
boundary and added to the simulation as additional pseudo-particles.
After this is done for all reflecting boundaries, we calculate all the
time derivatives in the usual manner treating all particles in an
equal way. When all time derivatives are calculated we delete all the
pseudo-particles.  Only the ``real'' particles are then advanced in
time. This simple algorithm allows us to implement reflecting boundary
conditions using only a small number (typically 10\% of the total
number of particles for the three boundaries we use) of additional
particles.

The use of SPH requires adding some artificial viscosity in order to
resolve shocks. We use the standard artificial viscosity
\citep[e.g.][]{mon_rev,benz_rev} consisting of a term analogous to
bulk viscosity and a Von Neuman-Richtmyer artificial viscosity term.
For the time integration we used a second order Runge-Kutta integrator
with an adaptive stepsize control.

\section{Results}
\label{sec:Res}

\begin{figure}
  \begin{center}
    \plotone{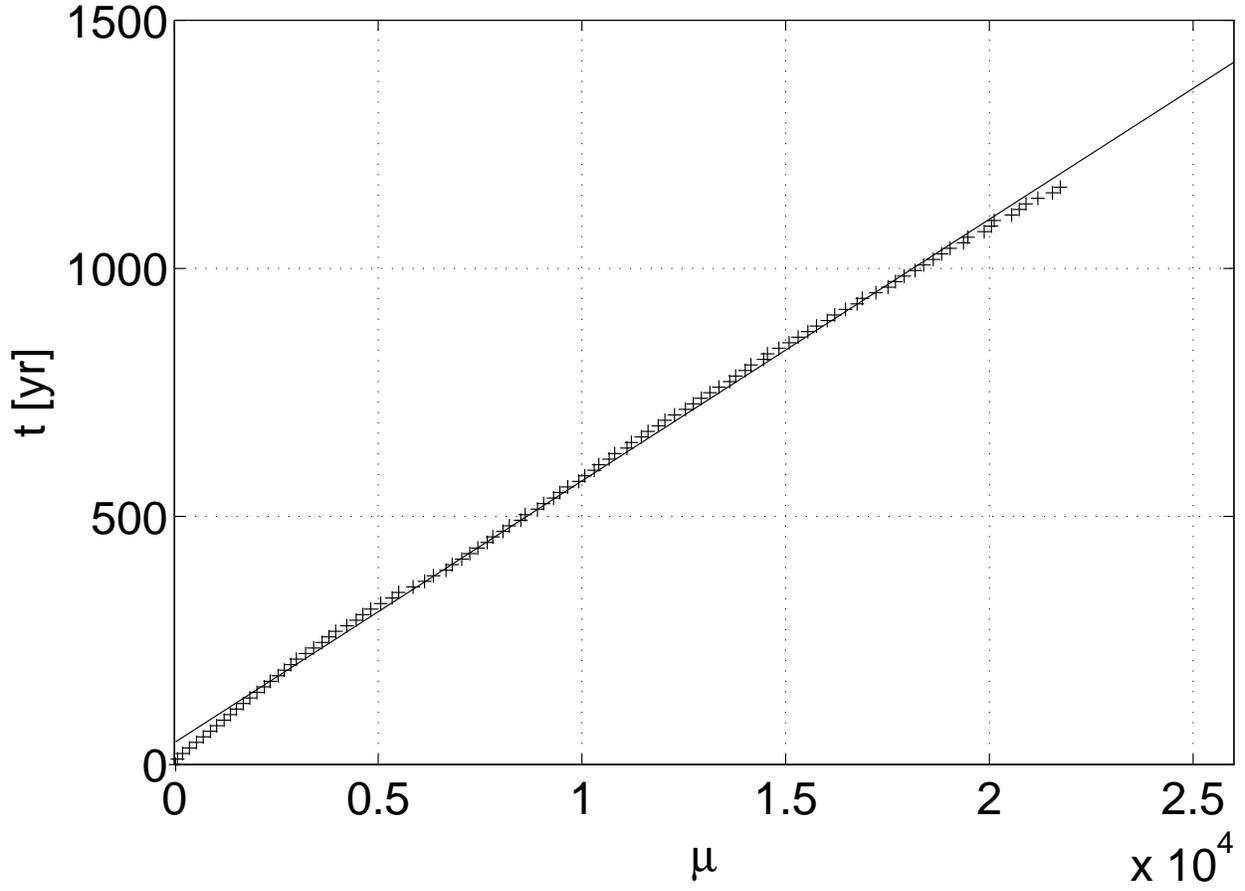}
    \caption{
      Time as a function of $\mu$. The linear relation between time
      and $\mu$ is $t\sim 0.046\mu(E_{51}/\ri)^{-1/3} \unts{yr}$. In
      this and all subsequent figures the results are presented for an
      initial energy of $10^{51}\unts{erg}$ and ISM density of
      $10^{-24}\unts{gm\,cm^{-3}}$.  }
    \label{fig:mu}
  \end{center}
\end{figure}

We choose the initial velocity to be in the range $0.22c$ to $0.33c$,
making relativistic effects small. Equation (\ref{eq:const}) leaves
the freedom of choosing two out of the three parameters $E_0$, $R_0$
and $\ri$, the initial energy, distance and ISM density respectively.
In presenting the results we choose $E_0$ and $\rho_0$.  To
parameterize the evolution of the remnant we utilize the fact that
mass scales linearly with the initial energy and we define the
dimensionless parameter $\mu=Mc^2/E_0$ where $M$ is the accumulated
shell mass. We define $M$ as all mass with density above $2\rho_0$
corresponding to the shocked material of the shell and to the
accumulated ISM mass. All subsequent results are presented as
functions of $\mu$. Our simulation begins approximately 2 years after
the burst, when $\mu\sim 18$. In Figure \ref{fig:mu}, we show the
linear scaling of time with $\mu$.  The scaling relation is $t\sim
0.053\mu(E_{51}/\ri)^{-1/3} \unts{yr}$ which is close to the
Sedov-Taylor \citep{Sedov59} blast wave result of $t\propto
R^{5/2}\propto\mu^{5/6}$ where $R$ is the radius of the blast wave.

\begin{figure}[htbp]
  \begin{center}
    \plottwo{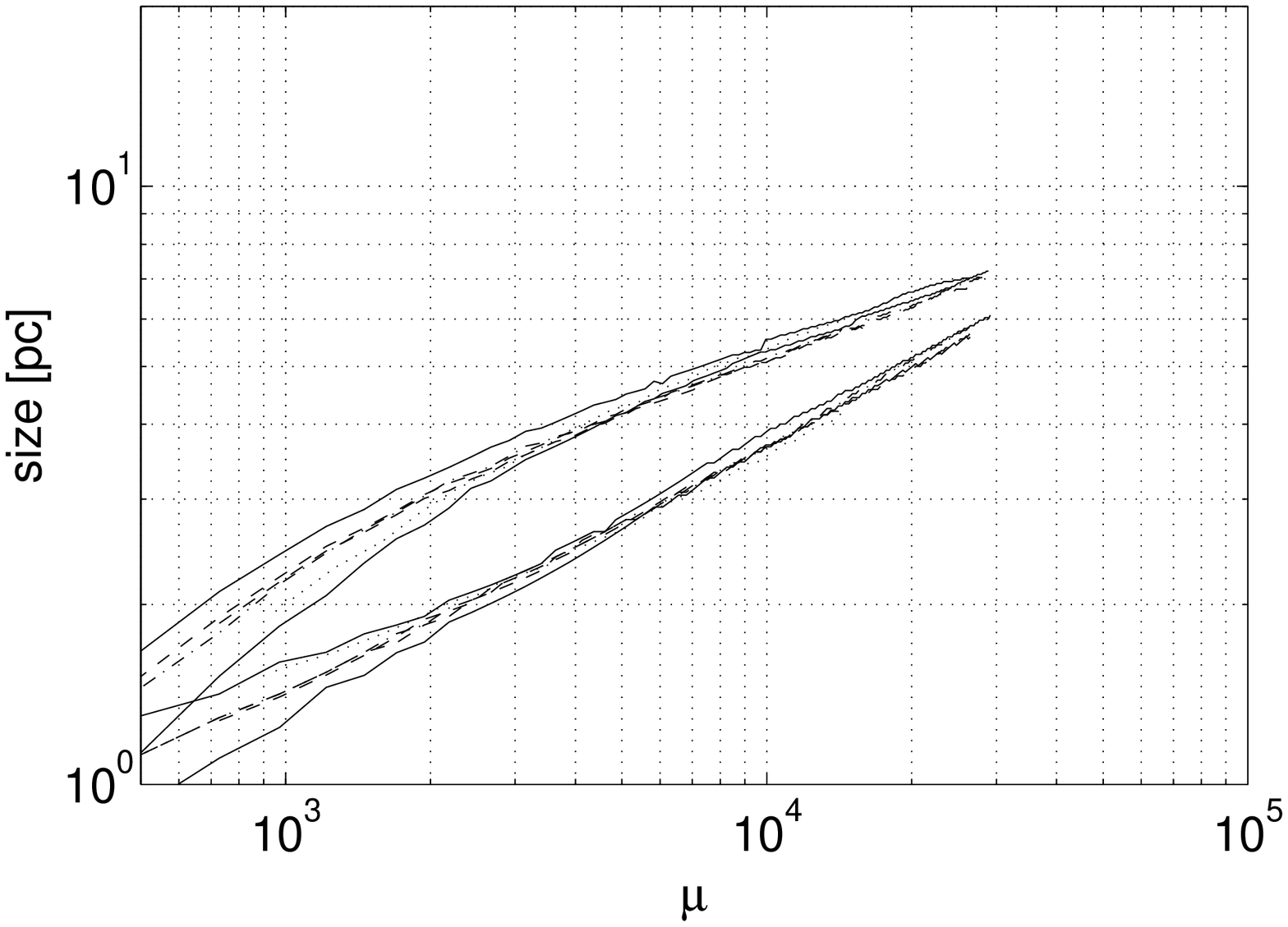}{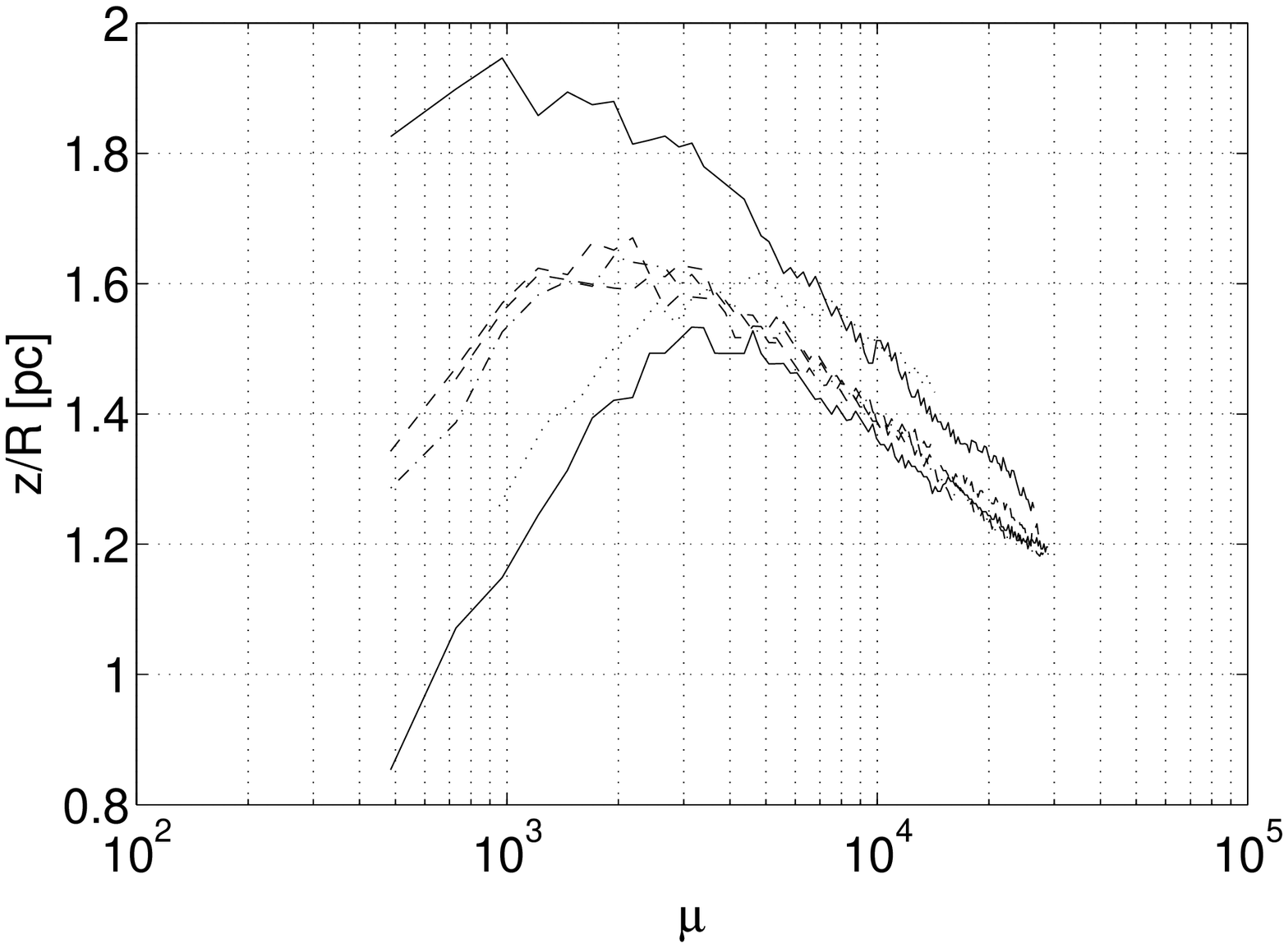}
    \caption{Comparison of the various runs showing their
      similarities. (a) The $z$ and
      $r_{xy}$ positions of the shocks (the $z$ positions are the
      higher lines). (b) The ratio $z/r_{xy}$ for all the runs. }
    \label{fig:posall}
  \end{center}
\end{figure}

A bow shock forms as each blob collides with the ISM. The shock also
propagates in the direction perpendicular to the blob's velocity and
over time, backwards. The expected morphology of the remnant will
therefore be of two expanding shells which will eventually join,
producing yet another shock. At late times the shells will merge and
become a single spherical shell. We made several runs, differing in
the initial density, shape, initial velocity, angular width and the
numerical coefficient in Eq. (\ref{eq:const}) of the ejecta as
summarized in Table~\ref{tab:runs}. In five runs the initial ejecta
was shaped like a disk and in one case it was shaped like a sphere. As
we show in Fig.~\ref{fig:posall}, for $\mu\gtrsim 5\times 10^3$ the
$z$ positions of the shock and its maximal radius in the $x-y$ plane,
$r_{xy}$ are indeed unaffected by these variations in initial
conditions. The $\mu$ at which the shells touch depends on the exact
initial conditions, but all other events (such as the $\mu$ at which
the remnant becomes spherical) are unaffected. Therefore in the
following discussions we will consider only one of the runs.

\begin{figure}
  \begin{center}
    \plotone{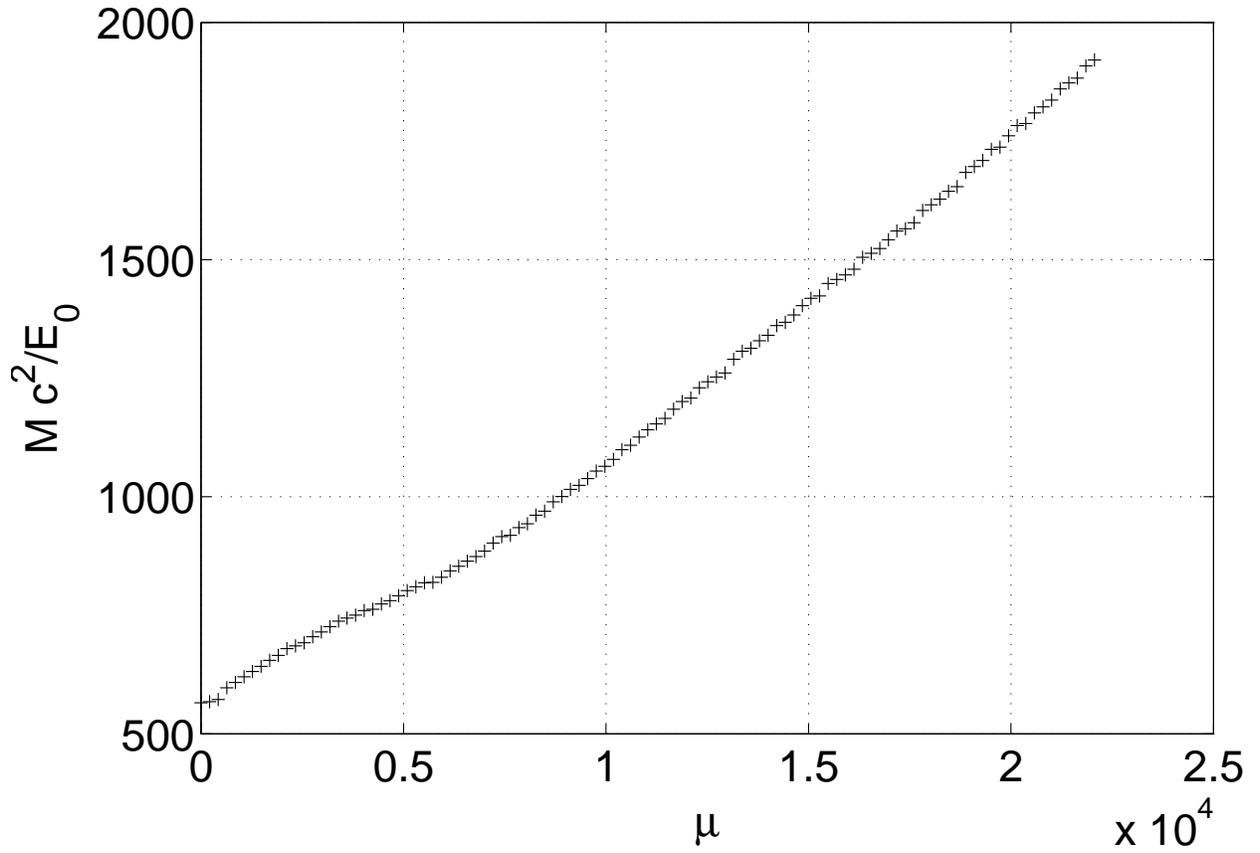}
    \caption{
      Mass inside the shell as a function of $\mu$.}
    \label{fig:mass}
  \end{center}
\end{figure}

\begin{figure}
  \begin{center}
    \plotone{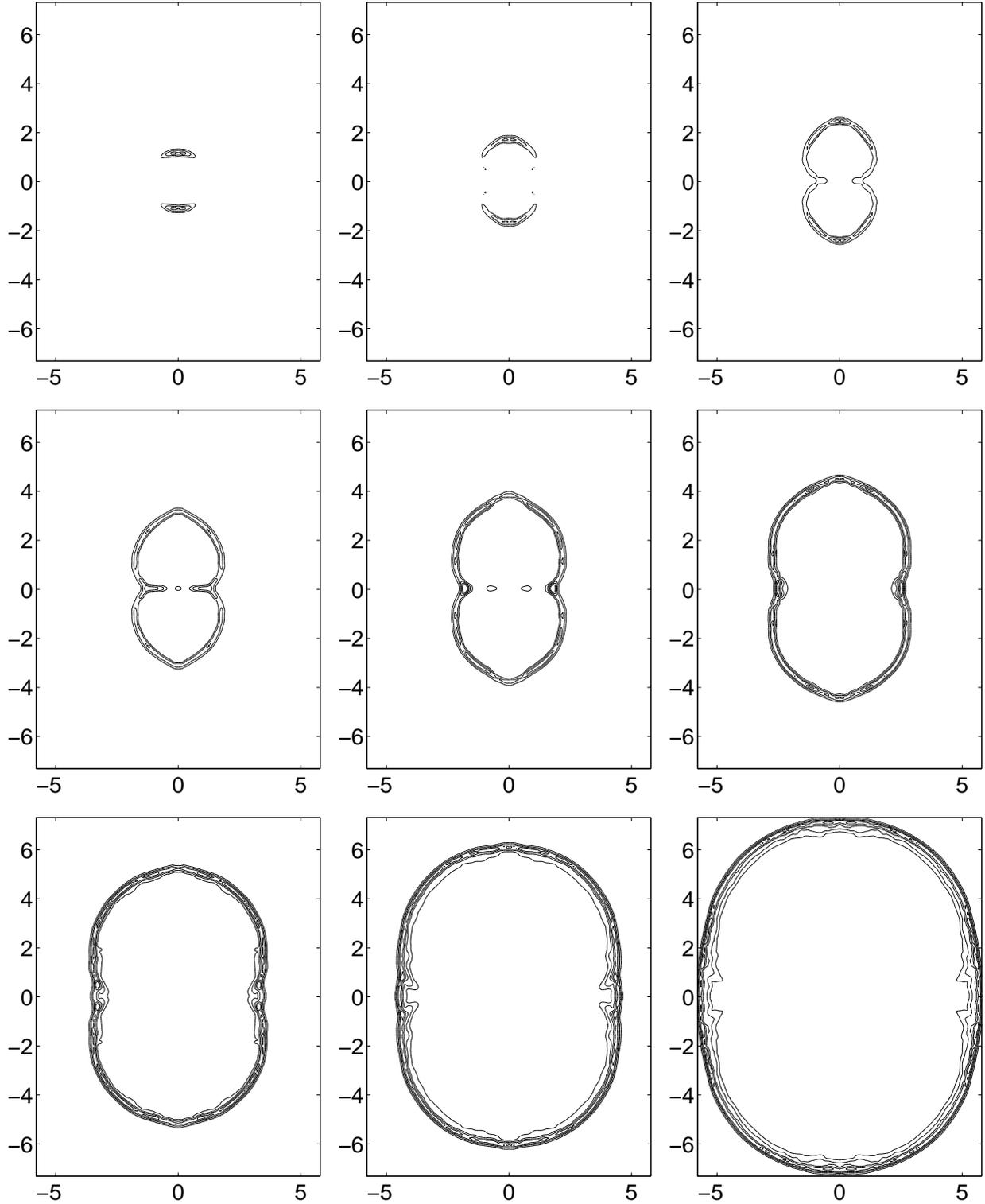}
    \caption{
      Density contours. The length scale is in pc. Contours are equal
      spaced at $1.5\rho_0,\,2\rho_0,\,\ldots,\,3.5\rho_0$. The images
      are at $\mu=2.1\times10^2,\,4.2\times10^2,\,8.5\times10^2,\,
      1.5\times10^3,\,2.5\times10^3,\,4.5\times10^3,\,
      7.6\times10^3,\,1.3\times10^4,2.2\times10^4$ (left to right, top
      to bottom)}
    \label{fig:dens_cont}
  \end{center}
\end{figure}

\begin{figure}
  \begin{center}
    \plotone{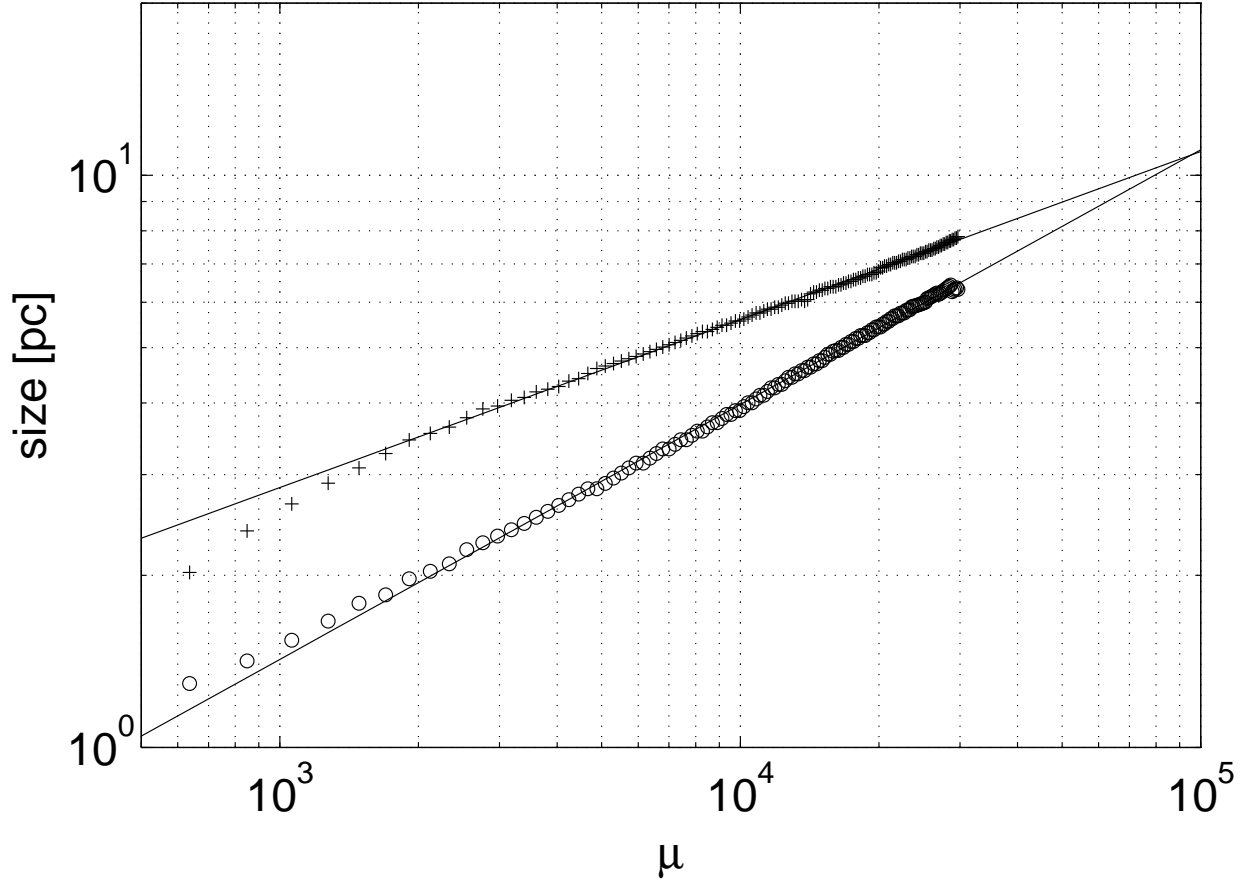}
    \caption{log-log plot of the $z$ position of the blobs (crosses)
      and the maximum radius in the $x-y$ plain of the shock $r_{xy}$
      (circles) as a functions of $\mu$. The solid lines are the best
      fit power laws for $\mu>2200$ which are $z\propto \mu^{0.29}$
      and $r_{xy}\propto \mu^{0.45}$.  The lines are extrapolated to
      the point where they cross at $\mu\sim 9.6\times 10^4$ }
    \label{fig:pos}
  \end{center}
\end{figure}

\begin{figure}[htbp]
  \begin{center}
    \plotone{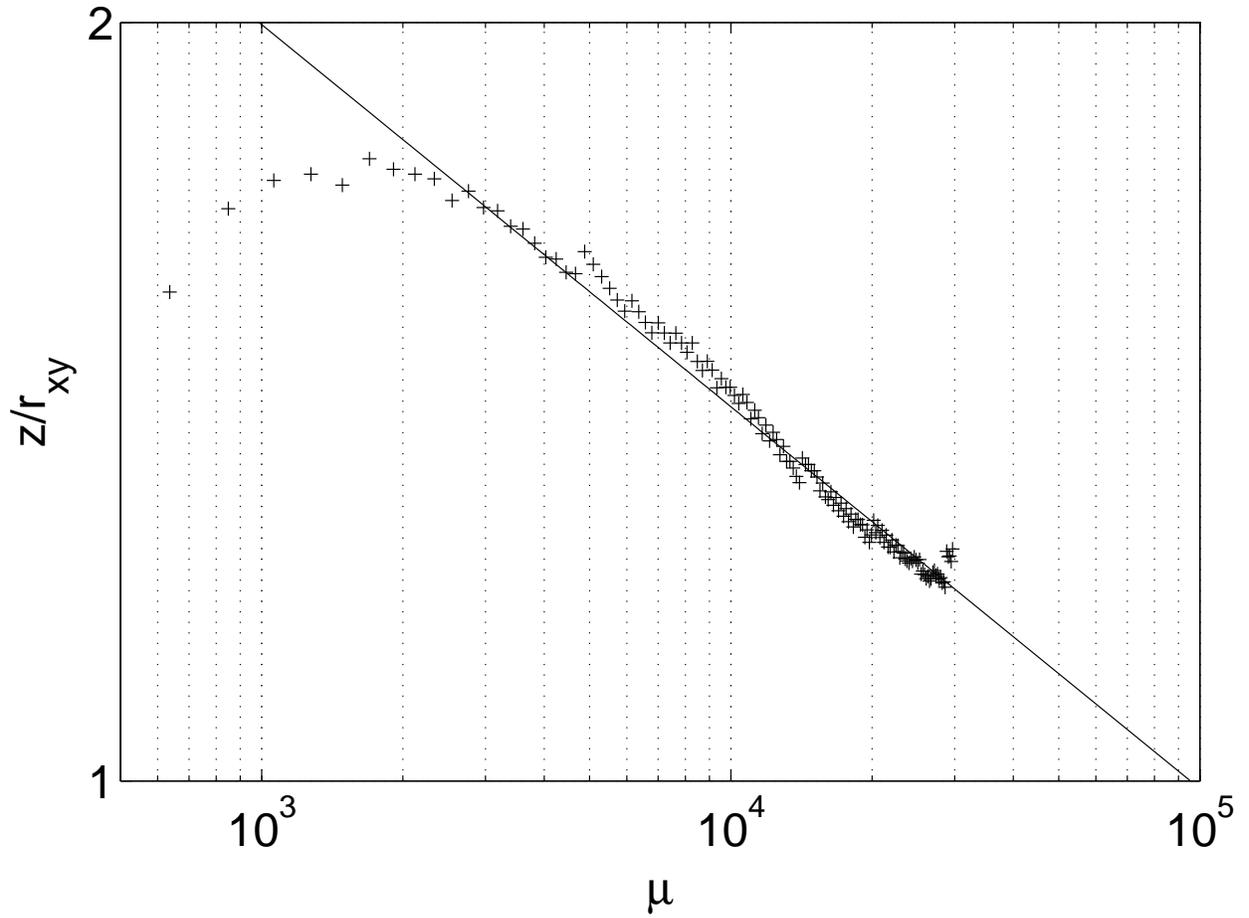}
    \caption{The ratio between the radius $r_{xy}$ of the shock and the $z$
      position of the shock. The solid line is the best fit power law
      $\mu^{-0.15}$. The ratio will reach a value of 1 at $\mu\sim
      9.6\times 10^4\unts{yr}$}
    \label{fig:ratio}
  \end{center}
\end{figure}

The mass inside the shell, defined as all mass with density below
$\rho_0/2$ also evolves almost linearly with $\mu$
(Fig.~\ref{fig:mass}) as $0.06(E_0/c^2)\mu$. In
Fig.~\ref{fig:dens_cont}, we show the density contours as a function
of $\mu$.  The two shells touch and a shock forms along the equatorial
plane between them at $t\sim 50-260(E_{51}/\ri)^{-1/3}\unts{yr}$ when
$\mu = 1-5\times 10^3$ depending on the initial conditions.  The
maximal $z$ position of the shock and $r_{xy}$ can be fitted with a
power law as shown in Fig.~\ref{fig:pos}. The scaling relation for
$r_{xy}$ is $r_{xy} = 0.07\mu^{0.45}(E_{51}/\ri)^{1/3}
\unts{pc}\propto t^{0.45}$. The effective radius of the shell,
$(r_{xy}^2z)^{1/3}$, scales as $t^{0.4}$, exactly the result expected
for a Sedov-Taylor blast wave.  The ratio $z_{max}/r_{xy}$ decreases
in time staying always between 1 and 2 (see Fig~\ref{fig:ratio}).
Extrapolating we see that this ratio reaches a value of 1 at $\mu\sim
9.6\times 10^4$. At this time the shock has a spherical shape with
$z=r_{xy}\sim 12(E_{51}/\ri)^{1/3} \unts{pc}$.  Even then the shock
will not be completely spherically symmetric as there would still be a
ring of shocks around the ``equator'' where the shells have collided.

\begin{figure}
  \begin{center}
    \plotone{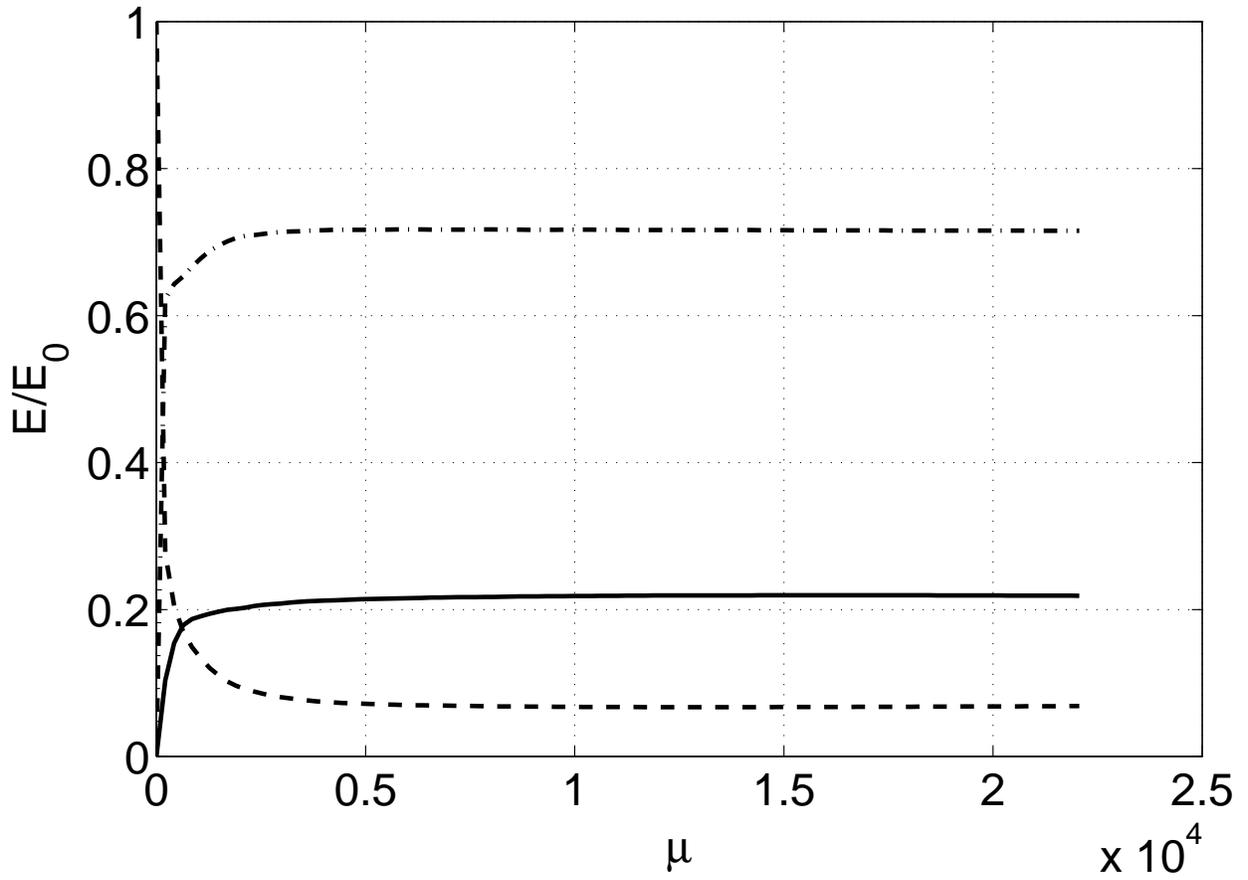}
    \caption{plot of total kinetic energy in the $z$
      direction (dashed line), total kinetic energy in the $x-y$
      direction (solid line) and total internal energy (dash-dotted
      line). The energies are scaled by the initial kinetic energy
      $E_0$.}
    \label{fig:energy}
  \end{center}
\end{figure}

During the whole run, total energy is conserved to within 1\%. In
Fig.~\ref{fig:energy} we show $E_z$ the total kinetic energy in the
$z$ direction, $E_{xy}$ the total kinetic energy in the $x-y$ plane
and $E_i$ the total internal energy in the simulated volume.  As the
blobs interact with the ISM $E_i$ and $E_{xy}$ increase at the expense
of $E_z$. $E_{xy}$ increases until $\mu\sim 5\times10^3$ and then it remains
constant at $0.22E_0$. $E_i$ increases by a factor of 200 to a value
of $0.72E_0$, most of the increase occurs before $\mu\sim2500$. In the
final configuration ($\mu\sim2\times10^4$) 72\% of the energy is in
internal energy, 22\% is in kinetic energy in the $x-y$ plane and only
6\% remains in the kinetic energy in the $z$ direction.

\begin{figure}
  \begin{center}
    \includegraphics[width=12cm]{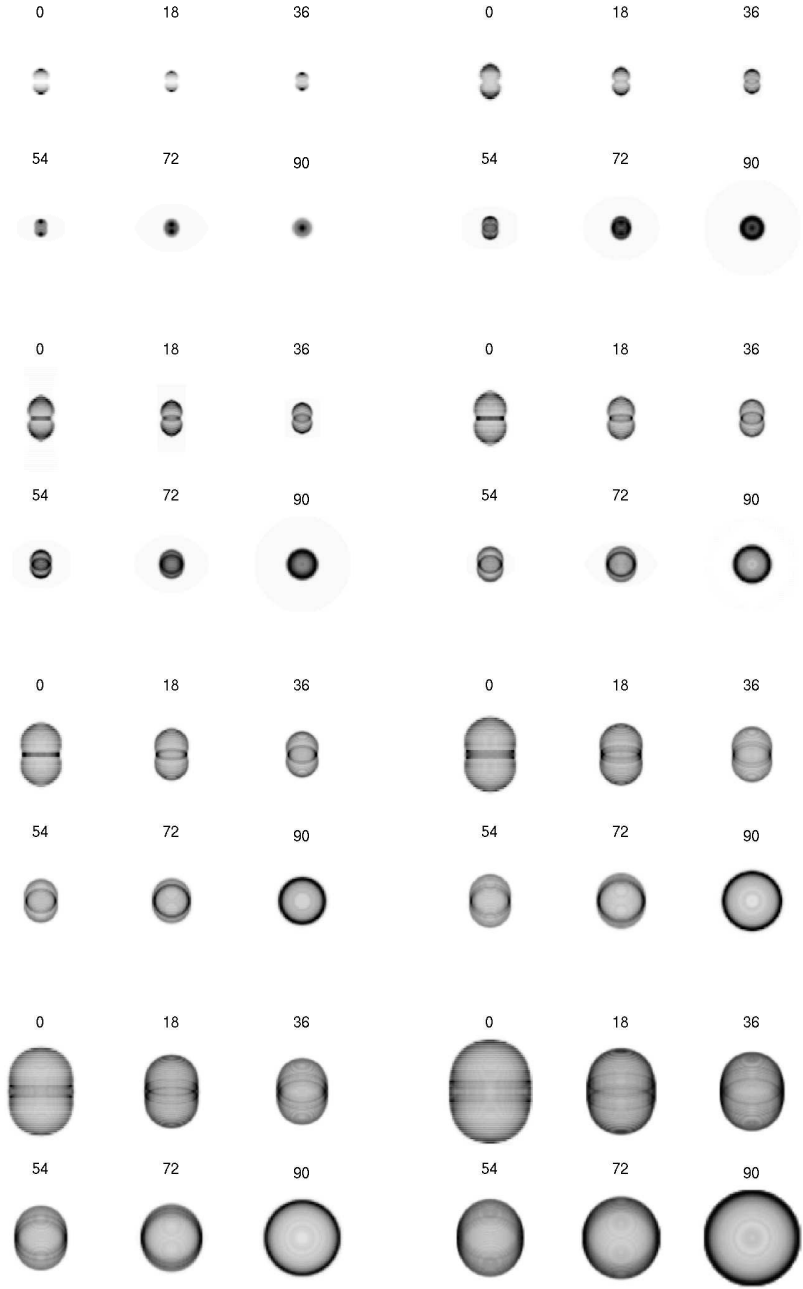}
    \caption{Images of the remnant, bremsstrahlung emission. The
      number above each image is the angle of inclination in degrees.
      The images are shown at the same $\mu$ as the last 8 panels of
      figure~\ref{fig:dens_cont}.  }
    \label{fig:images_brem}
  \end{center}
\end{figure}

\begin{figure}
  \begin{center}
    \includegraphics[width=12cm]{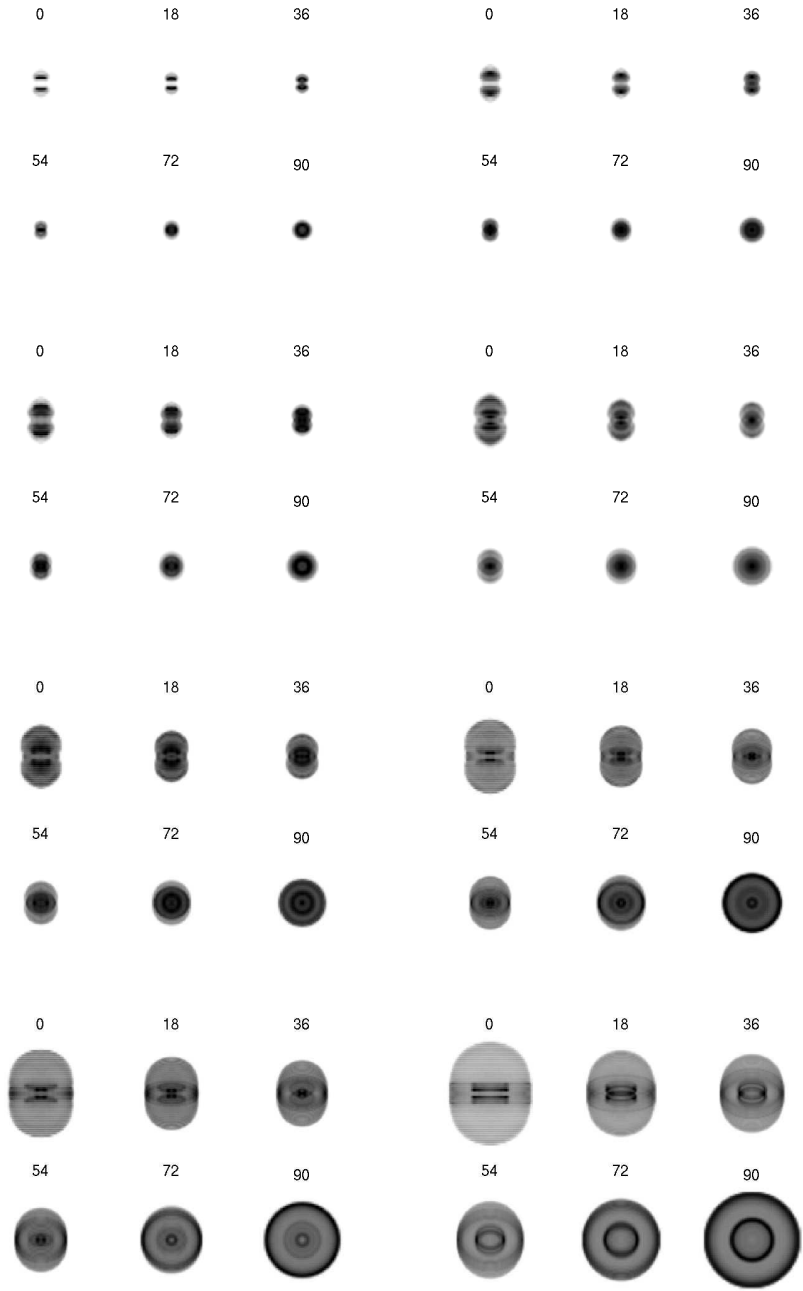}
    \caption{Images of the remnant, synchrotron emission. The $\mu$ are 
      the same as the last 8 panels in figure \ref{fig:images_brem}}
    \label{fig:images_sync}
  \end{center}
\end{figure}

Figures \ref{fig:images_brem} and \ref{fig:images_sync} depict the
images of the remnant as a function of time and angles of inclination.
We show images due to bremsstrahlung emission and synchrotron
emission. The images are constructed assuming that all the gas is
optically thin in the relevant frequencies. The bremsstrahlung
luminosity (Fig. \ref{fig:images_brem}) was calculated assuming that
the volume emissivity is proportional to $\rho^2\varepsilon^{1/2}$
\citep{lang80}. In calculating the synchrotron emissivity (Fig.
\ref{fig:images_sync}) we assumed that both the magnetic field energy
density and the energy of the relativistic electrons are proportional
to the internal energy density of the gas with the proportionality
factors $\epsilon_B$ and $\epsilon_e$ respectively. We further assume
that the relativistic electron number density is a power law in
energy. Under these assumptions the volume emissivity is proportional
to $\rho^2\varepsilon^2$ \citep[e.g.][]{shu91}. In the late images
there are two bright circles at the lines where the colliding blobs
forming a hot shocked region.

\section{Discussion}

We follow the hydrodynamic evolution of two blobs colliding with the
ISM. This scenario is a model for the intermediate time behavior of
matter ejected from a central engine producing a GRB. Our results can
be rescaled to fit a variety of initial energies and ISM densities.
The two shells touch and a shock forms along the equatorial plane
between them at $t\sim 50-160(E_{51}/\ri)^{-1/3}\unts{yr}$ when $\mu =
1-5\times 10^3$. We show that the late time remnant is insensitive
to the exact initial morphology, angular width and density of the
ejecta. Although initially the remnant may be highly non-spherical,
the ratio between its height and radius will approach unity and it
will eventually become spherical in shape after a time of $\sim
5\times 10^3E_{51}^{1/3}\ri^{-1/3} \unts{yr}$ when $\mu\gtrsim
9.6\times 10^4$. After this time it will be difficult to distinguish a
GRB remnant from a SNR using the morphology.  The expected number of
non-spherical GRB remnants is, therefore,
$5\times10^{-4}f_b^{-1}E_{51}^{1/3}\ri^{-1/3}$ per galaxy (using the
observed present GRB rate \citep{schmidt99} of
$10^{-7}\unts{yr^{-1}\,gal^{-1}}$).

The results of \cite{gaensler98} show a tendency for the bilateral
axis of the non-spherical SNRs to be aligned with the galactic plane.
This presents strong evidence in favor of an extrinsic model for the
origin of non-spherical SNRs. Using our results, we can propose an
alternative explanation to the origin of some of the highly
non-spherical SNRs and \HI~ shells. Instead of assuming an {\em
  extrinsic} model, namely spherical energy deposition into a non
isotropic medium we propose an {\em intrinsic} model: non spherical
energy deposition into an isotropic medium. Our GRB remnant model can
explain non-spherical SNRs with a two shell morphology provided that
$\mu\lesssim 9.6\times 10^4$. One extreme and well studied case, the
non-spherical SNR \dem~\citep{Williamsetal97}, has a distinct double
shell morphology. The external model clearly fails here. However the
ratio $\mu$ measured for \dem~is $\gtrsim 7.1\times10^5$ and a a GRB
remnant would already be spherical at this stage, suggesting that
\dem~ is not a GRB remnant.  This conclusion could be revised if for
some reason the initial blobs stay confined for a much longer period
(so that eq. 1 is not satisfied) or if the solution becomes radiative
before becoming spherical and the radiative cooling slow down the
evolution.

\acknowledgements 
We thank the anonymous referee for his helpful comments. This research
was supported by a ISRAEL-US BSF grant.

\begin{table}
  \begin{center}
    \begin{tabular}{clcc}
      run & shape & angular width & density [\ri]\\\tableline
      1 & disc   & 10/3 & 6/5 \\ 
      2 & disc   & 1    & 4/3 \\ 
      3 & disc   & 2    & 3   \\ 
      4 & disc   & 1    & 3   \\ 
      5 & disc   & 7/12 & 3   \\ 
      6 & sphere & 1/2  & 2      
    \end{tabular}
    \caption{Initial parameters for the different runs. Initial density 
      is in units of \ri.}
    \label{tab:runs}
  \end{center}
\end{table}

\end{document}